\documentstyle[eqsecnum,floats,preprint,aps,prd,epsf,rotate]{revtex}
\preprint{DAMTP R-97/61}
\date{\today}
\draft
\tighten
\begin{document}

\renewcommand{\arraystretch}{1.5}
\newcommand{\be}{\begin{equation}}
\newcommand{\ee}{\end{equation}}
\newcommand{\bea}{\begin{eqnarray}}
\newcommand{\eea}{\end{eqnarray}}
\def\Tr{\mathop{\rm Tr}\nolimits}
\def\sqr#1#2{{\vcenter{\hrule height.3pt
      \hbox{\vrule width.3pt height#2pt  \kern#1pt
         \vrule width.3pt}  \hrule height.3pt}}}
\def\square{\mathchoice{\sqr67\,}{\sqr67\,}\sqr{3}{3.5}\sqr{3}{3.5}}
\def\today{\ifcase\month\or
  January\or February\or March\or April\or May\or June\or July\or
  August\or September\or October\or November\or December\fi
  \space\number\day, \number\year}

\def\Bbb{\bf}


\title{Dynamic D8-branes in IIA string theory}

\author{A. Chamblin{$^1$} and M.J. Perry{$^1$}}

\address {\qquad \\ {$^1$}DAMTP\\
Silver Street\\
Cambridge, CB3 9EW, England
}

\maketitle

\begin{abstract}

In this paper we perform a detailed investigation of the Dirichlet
eight-brane of the Type IIA string theory, when the effects of
gravity are included.  In particular,  
consider what happens when one allows the
ten-form field strength $F_{10}$ to vary discontinuously across the worldvolume
of the brane.  Since the ten-form is constant on each side of the brane
($d*F_{10} = 0$), a variation in the bulk term $\int F_{10}*F_{10}$
gives rise to a net pressure acting on the surface of the brane.
This means that the infinite `planar' eight-brane is no longer a static
configuration with these boundary conditions.  Instead, a static configuration
is found only when the brane `compactifies' to the topology of an eight-sphere, $S^8$.
These spherical eight-branes are thus bubbles which form boundaries betweeen 
different phases of the massive Type IIA supergravity theory.
While these bubbles are generically unstable and will want to expand (or contract), 
we show that in certain cases there is a critical radius, $r_c$, at
which the (inward) tension of the brane is exactly counterbalanced by 
(outward) force exerted by the pressure terms.  Intuitively, these `compactified'
branes are just spherical bubbles where the effective cosmological constant
`jumps' by a discrete amount as you cross a brane worldsheet.
We point out that such configurations are very similar to Dirac's original
`membrane' model for the electron, whereby an electron is thought of as a
charged, spherical conducting surface such that $F^2 = 0$ inside the brane
and $F^2 \neq 0$ outside the brane (where $F = F_{\mu \nu}$ is the Maxwell
two-form field strength).  For this reason, we propose that these spherical
Dirichlet eight-branes can properly be known as `D(irac)-branes'.
We argue that these Dirac branes will be unstable to various semi-classical
decay processes.  We discuss the implications of such  processes for the
open strings which have endpoints on the eight-brane.

\end{abstract}


\section{Introduction}

In recent years, branes have emerged as an essential ingredient 
of string theory, which is our best hope for a `theory of everything'.
Most importantly, a M(agical) theory has recently come to light which has
taught us that all of string theory can be understood in a single framework,
unified by the force of duality symmetries.  This `M-theory' 
is already quantized, in the sense that it has no coupling constant and
therefore no classical limit.  Of course, it does admit a low-energy
limit, namely eleven dimensional supergravity.  The bosonic sector
of eleven dimensional supergravity contains just two fundamental fields, the 
graviton and a three-form potential, and it follows
that there are two basic p-branes which solve the equations of motion:
an (electric) 2-brane and a (magnetic) 5-brane.  One can test what M-theory
has to say about the world around us by compactifying combinations of these
extended objects to lower dimensional configurations.  For example, all of 
the p-branes of ten dimensional (Type IIA) supergravity theory can be 
obtained from either the membrane or the fivebrane of eleven dimensional
supergravity by dimensional reduction \cite{paul}.

Perhaps the most important thing which string duality teaches us is that in order
to have a consistent string theory we have to include objects known as
`D-branes'.  These D-branes, which are just hyperplanes where open strings
are allowed to end, were shown by Polchinski \cite{joe} to be the carriers
of ten dimensional Ramond-Ramond (R-R) charge (i.e., D-branes are the
solitonic p-branes of Type II supergravity which carry R-R charge).  
T-duality requires the existence
of these D-branes \cite{dai}, and once stringy effects are taken into account,
the D-branes become dynamical objects.  Indeed, the collective coordinates
for the transverse fluctuations of a D-brane are just the massless open string
excitations that live on the brane worldvolume.

Once you allow for the existence of these D-branes, where open strings can end,
it is natural to start asking questions about their gross kinematical and/or
dynamical properties.  For example, is there any constraint on the topology of
these branes?  What happens when we push a brane away from extremality?
Can these branes `rip', or tear, by some semi-classical decay process
analagous to the decay of cosmic strings (by black hole pair creation)?
In this paper, we point out that it is possible to have `localized' branes,
of spherical topology, and that these compact branes will indeed want to decay
through a semiclassical tunneling process rather similar to the mode
described in \cite{shawn1}.  

Throughout this paper we are working in
signature $(-, +, +, ... , +)$.  Our convention for the sign of the curvature is
\[
{\nabla}_{a}{\nabla}_{b}u_{c} - {\nabla}_{b}{\nabla}_{a}u_{c} = R_{abcd}u^{d}
\]

Before proceeding with the principal construction, it is useful if we first
recall some basic facts about D-branes \cite{larus}.  To begin, let us focus
just on the bosonic terms in the effective field 
theory actions which arise from the IIA
and IIB string theories.  These field theories are of course just the
type IIA and type IIB supergravity theories.  In the Neveu-Schwarz-Neveu-Schwarz(NS-NS)
sector these two theories have identical field content, consisting
of a metric tensor $g_{{i}{j}}$, an antisymmetric rank two tensor
potential $b_{{i}{j}}$ and a scalar dilaton field $\phi$ (these fields
arise from the solutions of string theory where the worldsheet fermions have
anti-periodic `Neveu-Schwarz' boundary conditions in 
both the right and left moving sectors of a closed string, and so that is why 
they are called NS-NS).  On the other hand, in the Ramond-Ramond (R-R) sector
the field content of the two theories is quite different.  Explicitly, the
bosonic sector for the effective action of IIA supergravity is given as
\begin{eqnarray}
S_{IIA} &=&  {1\over 2\kappa^2} \int d^{10}x \, \sqrt{-g}
\Bigl[e^{-2\phi}
\bigl(R+4(\nabla\phi)^2 \nonumber \\
&  & -{1\over 2\cdot 3!}H^2 \bigr)  
-\bigl({1\over 2{\cdot}2!}{F_{(2)}}^2 
+ {1\over 2{\cdot}4!}{F_{(4)}}^2 \bigr) \Bigr] \nonumber \\
&  & -{1\over 4\kappa^2} \int F_{(4)}\wedge F_{(4)}\wedge b  \>,
\end{eqnarray}

\noindent whereas the effective action for IIB supergravity takes the form
\begin{eqnarray}
S_{IIB} &=&  {1\over 2\kappa^2} \int d^{10}x \, \sqrt{-g}
\Bigl[e^{-2\phi}
\bigl(R+4(\nabla\phi)^2 \nonumber \\
&  & -{1\over 2\cdot 3!}H^2 \bigr)  
-{1\over 2}(\nabla A_{(0)})^2 \nonumber \\
&  & -{1\over 2\cdot 3!}(A_{(0)}B+F_{(3)})^2 \Bigr] \>, 
\end{eqnarray}

\noindent Here, $F_{(n)} = ndA_{(n-1)}$ are the antisymmetric 
$n$-form R-R field strengths, with
$(n-1)$-form potentials $A_{(n-1)}$, and $H = 3db$ is the antisymmetric
NS-NS three-form field strength.  The string coupling $g_s$ is given in terms of
the dilaton by the relation $g_s = e^{\phi}$.  Thus, in both of the above
actions the NS-NS sector is multiplied by a factor of ${g_s}^{-2}$; this means 
that NS-NS effects arise in ordinary string perturbation theory.  On the other
hand, the R-R terms are not multiplied by any such factor and so we know that
fundamental string states do not carry R-R gauge charges.  As we have already 
mentioned, the mystery of the R-R gauge fields was resolved when it was shown
that R-R charge was carried by the topological defects known as D-branes \cite{joe}.

Now these actions do not tell the whole story about D-brane configurations.
This is because they really only tell us about what life is like for massless
closed string states which exist ``off'' of the D-branes and how these states
couple to the D-branes.  For this reason, the supergravity effective actions
are called the `bulk' terms.  There should also be an effective description of 
life on the brane worldvolume.  Indeed, such an effective world-volume action
(denoted $S_{WV}$) does exist and it is characterized by a D(p)-brane tension,
$T_{p}$, and R-R charge density form $f_{p}$ as shown below:
\begin{equation}
S_{WV} = -T_p {\int} d^{p+1} \sigma e^{- {\phi}/2} 
\sqrt{\det (h_{\mu \nu} + b_{\mu \nu} + 2 \pi F_{\mu \nu})} - \nonumber \\
f_{(p)} {\int} d^{p+1} \sigma A_{(p+1)}
\end{equation}

\noindent Here, $\sigma$ denotes the coordinates tangent to the brane 
worldvolume, $h_{{\mu}{\nu}}$, $b_{{\mu}{\nu}}$ and $\phi$ are the 
pullback of the (respective) spacetime fields to the brane worldvolume.
$F_{\mu \nu}$ is an ordinary Maxwell gauge field which lives on the brane
worldvolume.  Again, $A_{(p+1)}$ is an antisymmetric $(p+1)$-form R-R potential
which couples naturally to the p-brane worldvolume (hence the appearance of
the R-R charge density $f_{(p)}$).  This form of the action is very useful
if one is interested in studying `light' branes, which have negligible
gravitational fields (so that you can ignore the `bulk' supergravity 
contributions).  Indeed, several people have made a lot of mileage
recently (\cite{gazz}, \cite{curt}) from the fact that the first term in
(1.3) is more or less the old Born-Infeld action of electrodynamics
(in the limit where you allow the Kalb-Ramond three-form and dilaton
to vanish).  In this limit a beautiful picture emerges, in which fundamental
strings ending on a D-brane appear, from the point of view of the
brane worldvolume theory, as Coulomb-like point particle solutions
of the Born-Infeld theory on the brane (Gibbons refers to all such solitonic
configurations in Born-Infeld theory as `BIons').  Similarly, you can also think of
an M2 brane ending on an M5 brane as a `vortex'-type BIon living on the
M5 brane worldvolume, and so on.  While these are certainly beautiful and
promising results, it is still of some interest to understand how strings and
branes in general will interact when they are `heavy', and the effects of
gravity are taken into account.  Presumably, addressing this problem will have
to involve somehow writing down an effective action which interpolates between
the world-volume and bulk terms as gravity is turned up.

In this paper, we consider a slightly simpler problem, namely, how the
world-volume and bulk terms `play off' of eachother, when you allow the
brane to gravitate.  What we find is that it is
possible to exactly cancel the `tension' term in the world-volume action with
pressure-like bulk terms in scenarios where the D-brane is a compact
spherical surface.  These spherical eight-branes are ten-dimensional,
gravitating analogues of Dirac's original membrane model for the electron,
and so we will refer to these spherical D8-branes as `Dirac branes'.
In order to properly understand this appellation, 
it is useful if we briefly recall
the basic facts about Dirac's idea.

In 1962 Dirac presented his membrane model for the electron; in this picture,
the electron is represented as a conducting sphere, with the property that 
inside the sphere the electromagnetic field vanishes ($F^2 = 0$) whereas
outside of the sphere the field strength is non-vanishing.  Thus in this scenario,
to quote Dirac, ``the electron may be pictured as a bubble in the 
electromagnetic field''.  Of course, the discontinuous jump in $F^2$ as 
one moves across the membrane boundary of the bubble means that there will
be a net pressure, which will cause the bubble to expand.  In order to 
counteract this pressure, Dirac assumed that the membrane had some tension
capable of holding the bubble in place.  Indeed, Dirac showed that for generic
values of the field parameters there would always be an `equilibrium radius'
for the bubble, at which the bubble would be perfectly static.
Later, both classical and quantum instabilities
were shown to exist for this model \cite{kuti}, although
we feel it is worth pointing out that supersymmetry may imply
that the quantum instability is not so bad.  This membrane
model, which may be regarded as the first suggestion that extended objects may
play an important role in our attempts to describe subatomic phenomena,
was clever but outside the bounds of experimental error: 
Certainly the mass ratio $m_{\mu}/m_{e} \sim 200$ was left
unexplained.

Dirac's motivations for introducing this idea were numerous.  On the one hand,
the idea of assigning a finite size to elementary particles in order to 
somehow regularize the total energy of the Coulomb field was an old one.
On the other hand, evidence had emerged that the muon and the electron were
very similar in many ways, and that perhaps the muon could be viewed as an
excited electron.  If the electron was a sphere, then perhaps the muon
could be a `wiggly' sphere.
 
In this paper, we will find gravitating spherical 
D-branes which are reminiscent
of the old Dirac membrane model.  As we have already said,
the simplest situation where this occurs involves the
eight-brane of Type IIA theory, and it is there that we now turn out attention.

\section{The D(8)-brane revisited}

The Dirichlet eight-brane of Type IIA couples to the ten-form field strength
$F_{10}$ of the R-R sector.  As Polchinski \cite{joe} pointed out, this
ten-form is not a dynamical variable; rather, it is just a constant field
which generates a uniform physical energy density which permeates space.
More explicitly, the bulk term generated by this field strength takes the 
form
\begin{equation}
S_{bulk} = -\frac{1}{2{\kappa}^2} \int d^{10}x \, \sqrt{-g} 
\frac{1}{2{\cdot}10!}{F_{(10)}}^2
\end{equation}

\noindent Since $F_{10}$ is constant, this bulk term corresponds to an
effective cosmological constant term.  This constant is precisely the mass
term of the massive IIA supergravity theory.  
The first supersymmetric $m \neq 0$ lagrangian for the IIA
theory was written down by Romans \cite{roman}.  In his theory, the mass
arises from a Higgs mechanism whereby the two-form `eats' the vector.  However,
as Howe, Lambert and West (HLW) \cite{neil} have recently pointed 
out there are actually three
distinct Higgs mechanisms, corresponding to whether the vector eats the scalar,
the two-form eats the vector or the three-form eats the two-form.  
In this way, they have
constructed a {\it new} massive IIA supergravity, which has the appealing
property that it can be obtained by compactification of eleven-dimensional Minkowski
space on a circle, with the introduction of a Wilson line.
In this paper, we will always be assuming that the four-form field strength
(coming from the eleven dimensional supergravity bosonic field sector) is turned
off.  With this caveat, we then obtain the below equations of motion in
ten dimensions for the HLW theory:
\begin{eqnarray}
R_{ab} - \frac{1}{2}g_{ab}R &=& -2(D_{a}D_{b}{\phi} - g_{ab}D^{2}{\phi} + 
g_{ab}(D\phi)^{2}) \nonumber \\
&  & + \frac{1}{2}(F^{ac}{F_{b}}^c - \frac{1}{4}g_{ab}F^2) -
18m(D_{(a}A_{b)} -g_{ab}D^{c}A_{c}) \nonumber \\
&  & - 36m^{2}(A_{a}A_{b} + 4g_{ab}A^2) - 12mA_{(a}{\partial}_{b)}{\phi} \nonumber \\
&  & - 30mg_{ab}A^{c}{\partial}_{c}{\phi} - 144m^{2}g_{ab}e^{-2\phi} + fermions  \>
\end{eqnarray}

\begin{equation}
D^{b}F_{ab} = 18mA_{b}{F_a}^{b} + 72m^{2}e^{-2\phi}A_{a} -
24me^{-2\phi}{\partial}_{a}{\phi}
\end{equation}

\begin{eqnarray}
6D^{2}{\phi} - 8(D\phi)^2 = -R + \frac{1}{4}F^2 + 360m^{2}e^{-2\phi}
+ 288m^{2}A^2 +96mA^{b}{\partial}_{b}{\phi} 
- 36mD^{b}A_b  \>
\end{eqnarray}

\noindent where $F_{ab} = {\partial}_{a}A_{b} - {\partial}_{b}A_{a}$ as usual.
The mass parameter $m$ determines the effective cosmological constant 
generated by the ten-form field strength.
These are the full bulk equations which we are free to play around with.
In this paper, we will be simplifying things by truncating these equations
of motion by setting
the vector field to zero:
\[
A_{(1)} = 0
\]
With this assumption, things simplify considerably.
Indeed, the only non-trivial remaining equations when the dust settles are
the Einstein equation:
\begin{equation}
R_{ab} = 360m^{2}e^{-2\phi}g_{ab}
\end{equation}

\noindent together with the `Maxwell' equation (2.3), which implies that
the dilaton $\phi$ is a constant.
In other words, if we turn off all of the fields in this theory
except for gravity, we simply recover de Sitter spacetime.  The effective
cosmological constant is then given explicitly in terms of the mass as
\[
{\Lambda} = 1296m^{2}e^{-2\phi}
\]

Because this theory admits such a remarkably
simple truncation, we will restrict our attention to this theory
in this paper.  Of course, everything which we will do here could in
principle also be done for the old Romans massive IIA theory.  However,
in the Romans theory
things are considerably more complicated; basically, the dilaton can
never be constant, and in fact the scalar curvature is identically
zero.  It is only in the interest of simplicity that we do not consider
the Romans theory further here.  

Our plan now is simple. Our aim is to find all possible eight-branes in this theory
since these are the objects that couple electrically to $F_{(10)}$. Away from the brane,
spacetime obeys the source-free equation (2.5). The principle of equivalence tells us
that all sources of energy and momentum gravitate. The brane is no exception to this rule,
and its energy-momentum tensor can be calculated from the action (1.3).
The brane itself is infinitely thin, and therefore its energy-momentum tensor
can be regarded as a distributional object concentrated entirely on the brane itself.
The physics of this situation is then rather similar to that of domain walls in the
so-called
\lq\lq thin wall" approximation.
Thus the spacetime energy-momentum tensor of the brane can be written as
\begin{equation}
T_{ab}=S_{ab} \delta(\eta)
\end{equation}
where the argument of the delta-function is the single coordinate transverse
to the brane evaluated in an normal coordinate system centered on the brane,
and $S_{ab}$ is termed the surface energy-momentum tensor.
\begin{equation}
S_{ab}=\sigma h_{ab}
\end{equation}
where $h_{ab}$ is the metric induced on the brane by its embedding into spacetime,
and $\sigma$ is the brane's energy density. The pullback of $h_{ab}$ to the 
world-volume
is just $h_{\mu\nu}$, and it is this metric which appears in the worldvolume
action for the brane. From equation (1.3), we see that the energy density
scales in terms of the brane tension as
\begin{equation}
\sigma = T_p e^{-{\phi}/2}={1\over 32}\pi^{-11/2}e^{-3{\phi}/2}
\end{equation}

\noindent Thus, away from the brane we simply need to understand the solutions
to equation (2.5).  In other words, the eight-brane divides spacetime up into
seperate domains.  As one moves from one domain to another, there will be 
conditions on how to match the spacetime in one region to 
the adjacent one.  These  conditions will depend
on the geometry and the energy density in the brane.


\newpage
\section{Brane compactification, spherical symmetry and Dirac branes} 

The fact that eight-branes divide ten-dimensional spacetime up into
domains reminds one of domain walls in cosmology.
As is well known, various cosmological models assume that a variety of
phase transitions took place in the early universe.  In such transitions,
symmetries which are only valid at high temperatures are broken as the
universe cools down.  In the common picture, `bubbles' of the new phase
are nucleated in regions of the old phase and may expand; if the rate of
production of these bubbles is not diluted by the rate of expansion of the
universe, the process of bubble nucleation, expansion and amalgamation
will continue until the universe is filled with new phase (with perhaps a few
bubbles of old phase `remnants' left over).  When the new phase fills the
entire universe, the transition is said to be complete.  One of the
principal conclusions of this work is that similar physics appears to 
occur in string theory.

The dynamical evolution of these bubbles, 
when the effects of gravity are included, has been studied by a number of
authors \cite{guth}, \cite{ber}, \cite{cvetic}.  These studies involve understanding
the Einstein equations when the source is a `thin shell', or domain wall.
In such situations, the spacetime has low differentiability and so one
has to regard the curvature as a distribution.
It was shown long ago \cite{israel} that the correct formalism for studying
such a problem involves constructing metric junction conditions for the
thin shells.  These junction conditions, commonly referred to as the
`Israel matching conditions', may be summarized as follows:
\begin{enumerate}
\item 
A domain wall hypersurface is totally umbilic; that is, the second
fundamental form $K_{ij}$ is proportional to the
induced metric $h_{ij}$ on the domain wall world-volume.
\item 
The  discontinuity in the second fundamental form on the domain wall hypersurface
is $[K_{ij}]_\pm = 4 \pi \sigma h_{ij}$, where $\sigma$ denotes the energy 
density of the domain wall.
\end{enumerate}

Thus, the energy density of an idealized domain wall measures the jump in the 
extrinsic curvature of surfaces parallel to the wall as you move through 
the wall.
We will use these conditions to do `cut-and-paste' constructions of 
eight-brane domain wall hypersurfaces.

Another point, worth bringing up, is that 
each side of these eight-branes will be regions where strings
are free to propagate in the bulk, and to attach themselves to the worldvolume
in the usual way, i.e., these branes will still be D-branes.  Presumably,
there will therefore be some minimal size for these objects (of order the
string length).  Any statements which we make about these objects, and how
they interact with strings, must therefore assume that they are bounded in
size in this simple way.

\subsection{Homogeneous and isotropic branes in the HLW theory}

We imagine the brane to be a surface sitting in spacetime. The simplest
geometry we can expect for it is to be a surface of constant curvature.
We therefore expect the brane to be spherical, planar or
hyperbolic.  Exterior to the brane, we expect the spacetime to 
reflect this geometrical property. 
We also suppose that the space-time orthogonal to the wall
is {\it static}. This comes from the
fact the observers co-moving with the wall would expect 
their situation to be time independent.
Furthermore, we assume that the directions parallel to the
brane worldsheet are homogeneous and isotropic.
Of course, this does not necessarily mean that the entire spacetime has to be
static,
it could well be that the orthogonal directions show some kind of 
cosomological expansion without violating these symmetry assumptions.
With these assumptions in mind, it follows
that the ten-dimensional metric takes the explicit form:
\begin{equation}
ds^2 = e^{2A(z)}(-dt^2 + dz^2 + f(t)^{2}[\frac{dr^2}{1 - kr^2} + r^{2}d{\Omega}_{7}])
\end{equation}

\noindent Here, $t$ denotes the time experienced by observers who move with
the wall, so that $A$ is a function of $z$ only and $f$ 
is a function of $t$ only (i.e., we have used boost invariance and so
$f$ measures the `scale' of the brane), $d{\Omega}_{7}$ denotes the metric
on a unit seven-sphere, and $k$ is the spatial curvature of the brane.
In other words, we have manifestly
decomposed the metric into a `normal to the brane' part (the $(t,z)$ sector) and a 
`parallel to the brane' part.  In the directions parallel to the
brane worldvolume, we have used our assumption
that the brane is isotropic and homogenous to impose key constraints
on the geometry.  Indeed, homogeneity and isotropy imply that the brane
worldvolume is an $(8 + 1)$-dimensional FLRW universe, and so that is why
we have written that down for the tangential part of the metric
(similar symmetry considerations were utilized in \cite{cvetic},
where similar configurations were studied in four dimensions).
For various reasons, we prefer to write the spatial part of the brane
metric in conformally flat form.  In other words, we introduce coordinate
$\rho$ and function $g(\rho)$ such that
\[
\frac{dr^2}{1 - kr^2} + r^{2}d{\Omega}_{7} = g(\rho)^2[d{\rho}^2 + 
{\rho}^2d{\Omega}_{7}]
\]

\noindent where $r$ and $\rho$ are related in the usual way:
\[
\int \frac{d\rho}{\rho} = \int \frac{dr}{r} (1 - kr^{2})^{-1/2}
\]

\noindent There are thus three basic cases to consider, corresponding to
the three types of FLRW universe:

{\noindent \bf Case 1, $k = 0$}\\

\noindent In this case, $k = 0$ and so the spatial section of the
brane is a flat plane.  For this case,
$r = \rho$ and $g(\rho) = 1$.

{\noindent \bf Case 2, $k > 0$}\\

\noindent In this case the brane has constant positive curvature, and so the
spatial section is an eight-sphere.  One calculates that 
\[
r = \frac{2(k)^{-1/2}{\rho}}{1 + {\rho}^{2}}
\]
\noindent whence
\[
g(\rho) = \frac{2(k)^{-1/2}}{1 + {\rho}^{2}}
\]

{\noindent \bf Case 3, $k < 0$}\\

\noindent In this case the brane has constant negative curvature, and so the
spatial section is
isometric to hyperbolic space.  Here, the transformation between $r$ and $\rho$
takes the form
\[
r = \frac{2(-k)^{-1/2}{\rho}}{1 - {\rho}^{2}}
\]
\noindent so that one finds
\[
g(\rho) = \frac{2(-k)^{-1/2}}{1 - {\rho}^{2}}
\]

\noindent With all of this in mind, it follows that the metric on each
side of the brane is given as
\begin{equation}
ds^2 = e^{2A(z)}(-dt^2 + dz^2 + 
f(t)^{2}g(\rho)^{2}[d{\rho}^2 + {\rho}^{2}d{\Omega}_{7}])
\end{equation}

\noindent In order to compute the curvature,
we introduce the orthonormal basis of one-forms $\{E^i: i=0,..., 9\}$
related to the coordinates by:
\[
E^0 = e^{A}dt,\ \ E^1 = e^{A}dz,\ \ E^i = e^{A}f(t)g(\rho)dx^i
\]

\noindent whence
\[
dt = e^{-A}E^{0},\ \ dz = e^{-A}E^{1},\ \ dx^{i} = \frac{e^{-A}}{gf}E^i
\]

\noindent A straightforward calculation yields the connection one-form
\begin{eqnarray}
{{\Gamma}^0}_i = \frac{{\dot f}}{f}e^{-A}E^i \nonumber \\
{{\Gamma}^0}_1 = A^{\prime}e^{-A}E^{0}  \nonumber \\
 \\
{{\Gamma}^1}_i = -A^{\prime}e^{-A}E^i \nonumber \\
{{\Gamma}^i}_j = e^{-A} \frac{{\partial}_{\rho}g}{fg^{2}{\rho}}(x^{j}E^{i} - 
x^{i}E^{j}) \nonumber 
\end{eqnarray}

\noindent where (for now), `dot' denotes differentiation relative to $t$, and
`prime' denotes differentiation relative to $z$.  
Proceeding, one obtains the non-zero components of the Ricci
tensor in the orthonormal frame, after using the specific functional form
for $g(\rho)$:
\begin{eqnarray}
R_{00} = e^{-2A}(A^{\prime\prime} 
+ 8{A^{\prime}}^{2} - 8\frac{{\ddot f}}{f})  \nonumber \\
R_{11} = e^{-2A}(-9A^{\prime\prime})  \\
R_{ij} = e^{-2A}(\frac{{\ddot f}}{f} +7\frac{{\dot f}^{2}}{f^2} + 7\frac{k}{f^2} 
- A^{\prime\prime} - 8{A^{\prime}}^{2}){\delta}_{ij}  \nonumber
\end{eqnarray}

\noindent From here we take the trace to obtain the Ricci scalar and so 
we construct the Einstein tensor $G_{\mu\nu}$.  From what we said earlier
about the massive HLW IIA theory, we are interested in solving the vacuum
Einstein equations with a cosmological constant, 
which take the form (in ten dimensions):
\begin{eqnarray}
e^{-2A}(A^{\prime\prime} + 8{A^{\prime}}^{2} - 8\frac{{\ddot f}}{f}) =
\frac{-1}{4}{\Lambda} \\
e^{-2A}(-9A^{\prime\prime}) = \frac{1}{4}{\Lambda} \\
e^{-2A}(\frac{{\ddot f}}{f} +7\frac{{\dot f}^2}{f^2} + 7\frac{k}{f^2}
- A^{\prime\prime} - 8{A^{\prime}}^{2}) = \frac{1}{4}{\Lambda}
\end{eqnarray}

Notice first that we have not specified the sign of the cosmological constant;
this is because we are interested in finding all possible solutions, before
restricting our consideration to the HLW sector (where the effective cosmological
constant has to be greater than or equal to zero).
The most obvious approach to solving these equations is to begin
with (3.6), and to try and solve for $A$.  Proceeding in this way one obtains
the integral 
\begin{equation}
\int \frac{dA}{(c^{2} - \frac{\Lambda}{36}e^{2A})^{1/2}} = {\pm}(z - z_{0})
\end{equation}

\noindent where $c$ and $z_{0}$ are some constants of integration.
There are three classes of solutions to this equation, corresponding
to the three possible signs of $\Lambda$.  We enumerate them as follows:

{\noindent \bf Class I, $\Lambda = 0$:}\\

In this case, to put it in the language of the massive supergravity theory,
we are looking for the behaviour of the overall conformal factor $e^{2A}$ in
a region where the theory is in a massless phase.  One easily integrates
(3.8) and finds that
\begin{equation}
e^{A} = e^{{\pm}c(z - z_{0})}
\end{equation}

\noindent so that $A^{\prime} = {\pm}c$, i.e., this would seem to imply that 
$c$ somehow measures the `acceleration' of the brane.  We will have more to say about
this point later.

{\noindent \bf Class II, $\Lambda > 0$:}\\

Here, we are looking at the conformal factor in a region of massive phase.
Again, one integrates and obtains for $A$ and finds that now
\begin{equation}
e^{A} = \frac{6c}{{\sqrt \Lambda}}\frac{1}{\cosh c(z - z_{0})}
\end{equation}

\noindent so that $A^{\prime} = -(c^2 - \frac{\Lambda}{36}e^{2A})^{1/2}$.

{\noindent \bf Class III, $\Lambda < 0$:}\\

This case corresponds to regions where the effective cosmological constant
is negative, or the mass is `imaginary', and as such it has nothing to do 
with the HLW theory.  Nevertheless, we include this case in the interest of
presenting a complete classification of the solutions.  Integrating with
negative $\Lambda$ one finds that
\begin{equation}
e^{A} = \frac{6c}{{\sqrt {- \Lambda}}}\frac{1}{\sinh c(z - z_{0})}
\end{equation}

\noindent Note that in this case $c^2$ can be less than zero, in which 
case $c$ is imaginary but the expression for $e^A$ is still valid.

These different classes summarize everything we need to know about the
behaviour of the conformal factor $e^{2A}$.  To understand
completely the geometry of these branes, therefore, we just need to
calculate the scale factor $f(t)$.  In order to do this, we combine
equations (3.5) and (3.7), and we use that fact that generically
$A^{\prime} = -(c^2 - \frac{\Lambda}{36}e^{2A})^{1/2}$,
in order to derive the identity
\begin{equation}
c^{2} = \frac{{\ddot f}}{f}
\end{equation}

\noindent Using this identity, one can work out all of the possibilities
for $f$.  These possibilities can be classified according to the sign of
$c^2$, as shown:

{\noindent \bf \underline{$c = 0$}:}\\

When $c = 0$, $f$ assumes the form
\[
f(t) = {\alpha}t + {\beta}
\]

\noindent so that $k$ satisfies $k = -{\alpha}^{2} < 0$.

{\noindent \bf \underline{$c^{2} > 0$}:}\\

When $c^{2} > 0$ the scale factor assumes the form
\[
f(t) = {\alpha}e^{ct} + {\beta}e^{-ct}
\]

\noindent so that the curvature is given as $k = 4{\alpha}{\beta}c^2$.

{\noindent \bf \underline{$c^{2} < 0$}:}\\

When $c^{2} < 0$, $f$ can be written as
\[
f(t) = {\alpha}\sinh (ct) + {\beta}\cosh (ct)
\]

\noindent where $k = ({\alpha}^{2} + {\beta}^{2})c^{2}$.

Thus, we see that there are numerous possibilities for homogeneous, isotropic
and boost invariant branes which bound regions which are solutions of the
vacuum Einstein equations.  We summarize all of these possibilities, 
in terms of the sign of the curvature of the brane, in order
to emphasize the fact that spherical branes are preferred when
${\Lambda} > 0$  as shown below:

\begin{center}
{\bf Spherical branes ($k > 0$)}
\end{center}

Here, as we have seen, $c^2 > 0$ and so $k = 4{\alpha}{\beta}c^2$, where
$\alpha$ and $\beta$ have the same sign and the scale factor $f(t)$
for the brane is given as
\begin{equation}
f(t) = {\alpha}e^{ct} + {\beta}e^{-ct}
\end{equation}

\noindent The overall conformal factor then depends on the sign of the
cosmological constant as shown below:
\newpage
\begin{eqnarray}
{\Lambda} = 0 ~{\Rightarrow}~ e^{A} = e^{-c(z - z_{0})} \\
{\Lambda} > 0 ~{\Rightarrow}~ e^{A} = 
\frac{6c}{\sqrt{\Lambda}}\mbox{sech}[c(z - z_{0})] \\
{\Lambda} < 0 ~{\Rightarrow}~ e^{A} = 
\frac{6c}{\sqrt{-\Lambda}}\mbox{cosech}[c(z - z_{0})] \>
\end{eqnarray}

\noindent Thus, we see that it is possible to have spherical eight-branes for
any sign of the cosmological constant.  When ${\Lambda} = 0$ these are just the
ten-dimensional version of the old four-dimensional Vilenkin-Ipser-Sikivie
domain walls \cite{vis}.  When ${\Lambda} < 0$ we recover the ten-dimensional
version of the non-extreme supergravity domain walls discussed by Cvetic
et al \cite{mir}.  When ${\Lambda} > 0$ we are left with two-sided
de Sitter-de Sitter (dS-dS) type eight-brane `bubbles'.  The energy density of these
branes is calculated (using the Israel matching conditions) to be
\begin{equation}
\sigma = \frac{2}{\kappa}\Bigl({\sqrt{{\pm}{l_1}^2 + c^2}} + 
{\sqrt{{\pm}{l_2}^2 + c^2}}\Bigr)
\end{equation}

\noindent where ${l_i}^2 = +\frac{{\Lambda}_i}{36}$ when ${\Lambda}_i < 0$
and ${l_i}^2 = -\frac{{\Lambda}_i}{36}$ when ${\Lambda}_i > 0$.  Thus, for
the dS-dS type eight-branes of the HLW theory, the 
vacuum energy density on each side of the 
branes is bounded from above in terms of the parameter $c^2$:
\[
{{\Lambda}_i} ~{\leq}~ 36c^2
\]

Because these spherical Dirichlet eight-branes are reminiscent, from a 
ten-dimensional point of view, of the old Dirac membrane model for the electron,
we shall henceforth refer to them as {\it Dirac} branes.

\begin{center}
{\bf Planar branes ($k = 0$)}
\end{center}

Here, there are two possibilities for the scale factor.
We must take ${c^2} ~{\geq}~ 0$ and hence either $\alpha$ or $\beta$ must vanish
(when $c^2 > 0$) or $\alpha$ must vanish (when $c^2 = 0$).  In the $c^2 > 0$ case,
the brane scale factor therefore takes the form
\begin{equation}
f(t) = {\alpha}e^{{\pm}ct}  
\end{equation}

\noindent The conformal factor in this case is then given as
\begin{eqnarray}
{\Lambda} = 0 ~{\Rightarrow}~ e^{A} = e^{-c(z - z_{0})} \\
{\Lambda} > 0 ~{\Rightarrow}~ e^{A} = 
\frac{6c}{\sqrt{\Lambda}}\mbox{sech}[c(z - z_{0})] \\
{\Lambda} < 0 ~{\Rightarrow}~ e^{A} = 
\frac{6c}{\sqrt{-\Lambda}}\mbox{cosech}[c(z - z_{0})] \>
\end{eqnarray}

\newpage
\noindent In the $c^2 = 0$ case the brane scale function reduces to a constant,
and the conformal factor is given as
\begin{eqnarray}
{\Lambda} = 0 ~{\Rightarrow}~ e^{A} = constant \\
{\Lambda} < 0 ~{\Rightarrow}~ e^{A} = \frac{6}{\sqrt{-\Lambda}}\frac{1}{(z - z_{0})} \>
\end{eqnarray}

\noindent There is no positive $\Lambda$ case when $c^2 = 0$.  Thus, in the
context of the HLW theory it {\it is} possible to have planar branes, as long
as they are not static (i.e., as long as the scale factor is going like an
exponential).  At first glance, this might seem strange since it implies that
it is possible to slice de Sitter space along planar timelike hypersurfaces.
In fact, it is possible to find totally umbilic, timelike hypersurfaces which
are isometric to Minkowski space in de Sitter spacetime, as was first
demonstrated in \cite{xxx}.

\begin{center}
{\bf Hyperbolic branes ($k < 0$)}
\end{center}

Here, there are three possibilities for the scale function, corresponding to the
three possible signs for $c^2$.  For the case $c^2 > 0$, $f$ has the form
\begin{equation}
f(t) = {\alpha}e^{ct} + {\beta}e^{-ct}
\end{equation}

\noindent where $\alpha$ and $\beta$ have the opposite sign, and the curvature
is given as $k = 4{\alpha}{\beta}c^2$.  The conformal factor is given as
\begin{eqnarray}
{\Lambda} = 0 ~{\Rightarrow}~ e^{A} = e^{-c(z - z_{0})} \\
{\Lambda} > 0 ~{\Rightarrow}~ e^{A} = 
\frac{6c}{\sqrt{\Lambda}}\mbox{sech}[c(z - z_{0})] \\
{\Lambda} < 0 ~{\Rightarrow}~ e^{A} = 
\frac{6c}{\sqrt{-\Lambda}}\mbox{cosech}[c(z - z_{0})] \>
\end{eqnarray}

When $c^2 = 0$, the scale factor has the form
\begin{equation}
f(t) = {\alpha}t + {\beta}
\end{equation}

\noindent where $k = -{\alpha}^2$, and the conformal factor is
\begin{eqnarray}
{\Lambda} = 0 ~{\Rightarrow}~ e^{A} = constant \\
{\Lambda} < 0 ~{\Rightarrow}~ e^{A} = \frac{6}{\sqrt{-\Lambda}}\frac{1}{(z - z_{0})} \>
\end{eqnarray}

\noindent Again, there is no positive $\Lambda$ case when $c^2 = 0$.
When $c^2 < 0$, the scale factor is
\begin{equation}
f(t) = {\alpha}\sinh (ct) + {\beta}\cosh (ct)
\end{equation}

\noindent where $k = ({\alpha}^{2} + {\beta}^{2})c^2$, and so there is only one
possibility for the conformal factor
\begin{equation}
e^{A} = {\sqrt{\frac{c^2}{\Lambda}}}\mbox{cosech}[c(z - z_{0})]
\end{equation}

Thus, we see that it is possible to have branes of constant negative spatial
curvature in the HLW theory, where the effective cosmological constant is
bounded from below by zero, again as long as the branes are not static.  
This is interesting, since given a hyperbolic brane $H^8$ it 
is of course always possible to find some 
freely acting discrete group $G$ such that the 
quotient ${H^8}/G$ is a brane with non-trivial topology.  In this way, one
should be able to construct D8-branes of arbitrary topology.  We will have
nothing more to say about this issue, or hyperbolic branes
in general, in this paper.

\subsection{A Dirac brane menagerie}

With all of the above in mind we proceed with an analysis of the spherically symmetric
solutions, which we dub the `Dirac branes'.  
To do this we first choose a gauge so that the spherically
symmetric ten-dimensional metric takes the form \cite{tang}, \cite{rob}:
\begin{equation}
ds^2 = -f(r)dt^2 + \frac{1}{f(r)}dr^2 + r^{2}d{{\Omega}_{8}}^2
\end{equation}

\noindent where f is a function of $r$ only.  The most general spherically
symmetric {\it vacuum} metric in $9 + 1$ dimensions will have the form (3.33), where
the potential $f$ is given as
\begin{equation}
f(r) = 1 - \frac{C}{r^7} - \frac{\Lambda}{36}r^2
\end{equation}

This is of course just the metric of Schwarzschild-de Sitter (SdS) spacetime,
where $\Lambda$ is the cosmological constant and $C$ is given in terms
of the ADM mass $M$ mass of the black hole as follows:
\[
C = (\frac{105G}{16{\pi}^3})M
\]

The reader may be wondering why we aren't including any $U(1)$ charges at 
this point.  Actually, we could have left $F_{{\mu}{\nu}}$ non-zero and 
considered the resulting Einstein-Maxwell-dilaton system of equations; we
begin with the $F = 0$ truncation for simplicity.  

Before we proceed with the construction of the equations of motion for the
brane worldvolume it is useful if we have a picture of the simplest Dirac
brane.
In the simplest scenario, a (static) Dirac brane is just a bubble of de Sitter space,
bounded by the asymptotically flat region exterior to a Schwarzschild black
hole.  That is
to say, we take one of our Schwarzschild solutions,
and cut along the timelike (static) worldsurface that satisfies the Israel matching
conditions.  Then we throw away the `interior' of the worldvolume of the brane,
and instead we paste in a timelike cylinder which we have cut out of 
de Sitter space.  Obviously, the Israel conditions can be satisfied at the 
junction hypersurface.  Furthermore, it is not hard to see that the brane
tension will be positive.  This construction is illustrated below:
\vspace*{0.5cm}

\epsfxsize=12cm
\epsfysize=10cm
\epsfbox{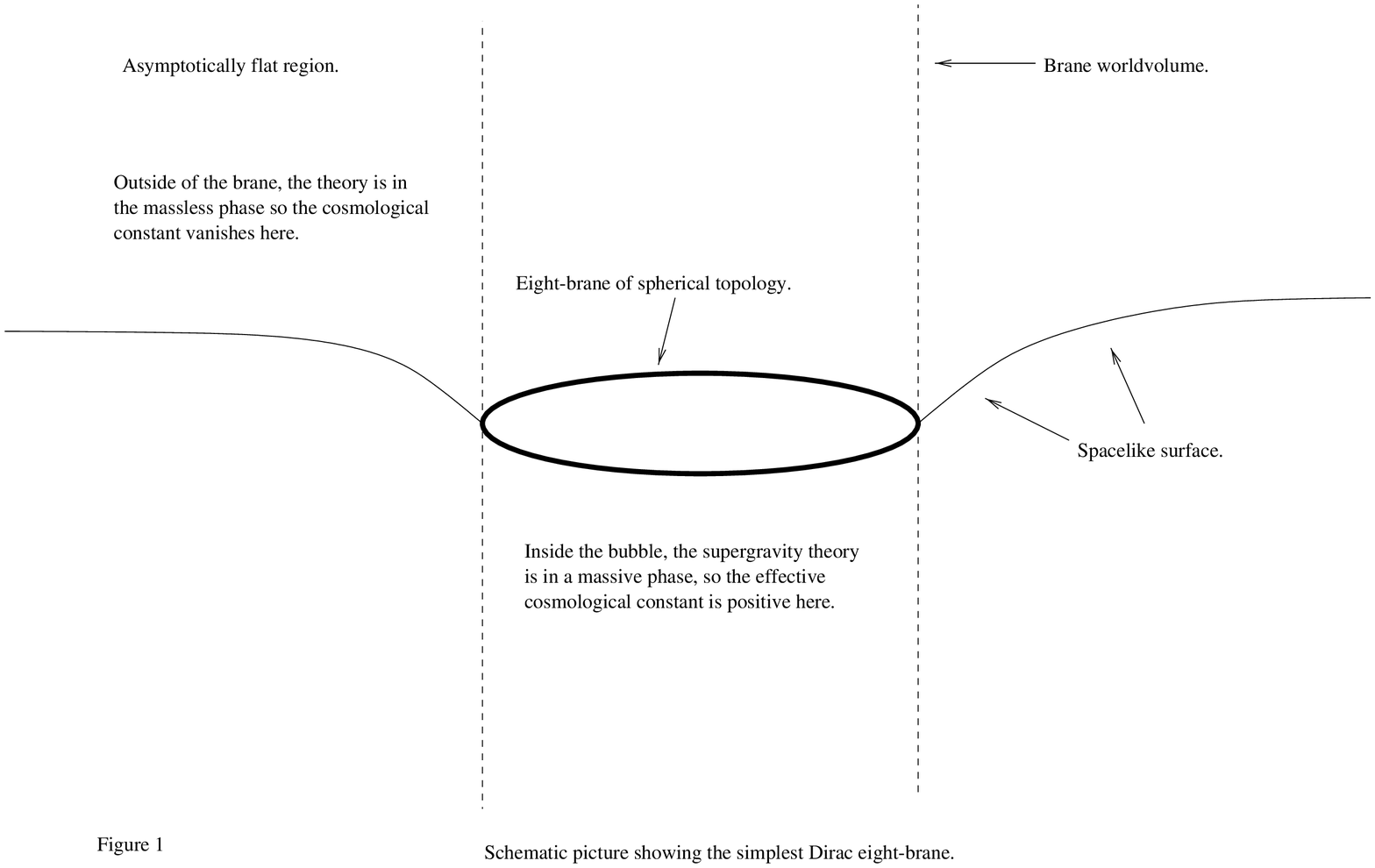}
\vspace*{0.5cm}

With this simple picture in mind we can now proceed with the equations of motion.
The equations of motion for a thin spherical shell, which is bounding a region of 
SdS from a region of a different SdS
spacetime, were worked out for four dimensions in \cite{ber}.  The analysis
developed there will go through throughout this paper, because the assumption
of spherical symmetry means that brane motion is described as a system in
$1 + 1$ dimensions.  Put more simply, we are solving for one unknown function,
$r(t)$, which describes the radius of the spherical brane at time $t$.
As always, we will let $\sigma$ denote the surface energy
density of the brane.  We will also distinguish between quantities which
exist `inside' of the brane, and those which are outside, by using subscripts.
Thus, $f_{in}$ denotes the metric function (3.34) inside of the brane,
and similarly $f_{out}$ denotes the same thing outside of the brane.
As we approach the brane from the inside, spheres parallel to the brane can
either be expanding or shrinking, and we introduce the symbol ${\epsilon}_{in}$
to distinguish these two cases.  That is to say, ${\epsilon}_{in} = +1$ if the 
spheres are growing as we move through the brane, ${\epsilon}_{in} = -1$ if 
the spheres are shrinking and similarly for ${\epsilon}_{out}$.  With all of
this in mind, the equation of motion for a spherical bubble bounding two different 
(vacuum) phases of the HLW supergravity theory is given as
\begin{equation}
{\epsilon}_{in}[{\dot r}^2 + f_{in}(r)]^{1/2} - 
{\epsilon}_{out}[{\dot r}^2 + f_{out}(r)]^{1/2} = \frac{{\kappa}\sigma}{2}r
\end{equation}

\noindent where ${\dot r} = f^{-1/2}{\partial}_{t}r$ denotes the derivative
relative to the time experienced by observers who co-move with the domain wall
and ${\kappa} = 8{\pi}G$ as always.
Any solution of this equation automatically describes an eight-brane worldsheet
which satisfies the Israel conditions.  Let us begin our investigation of 
equation (3.35) by assuming that both the inside and outside of the brane 
we turn off the Schwarzschild mass parameter, $M$.  Thus, inside of the brane
the metric potential is given as
\[
f_{in} = 1 - \frac{{\Lambda}_{in}}{36}r^2
\]
\noindent and similarly outside
\[
f_{out} = 1 - \frac{{\Lambda}_{out}}{36}r^2
\]

From equation (2.5) we know that the effective cosmological constants on each
side of the brane are given in terms of the characteristic masses of the supergravity
theory on each side by the relations 
${\Lambda}_{in} = 1296{m_{in}}^{2}e^{-2{\phi}_{in}}$ and 
${\Lambda}_{out} = 1296{m_{out}}^{2}e^{-2{\phi}_{out}}$.  
However, as we discussed above these
masses are basically the squares of the ten-form field strengths on each side
of the brane, and we know that R-R charge is quantized \cite{andy}.
In other words, the masses must differ by some discrete multiple of the 
fundamental unit of R-R charge.  

If we assume that each side of the eight-brane is a massive phase of the HLW
theory, then we are just pasting together two portions of 
two distinct ten-dimensional
de Sitter spacetimes.  Thus, the total spacetime is closed, i.e., spatial
sections are compact and homeomorphic to $S^9$ (with a `ridge' of curvature
running along where the location of the eight-brane).  
It follows that the distinction between `in' and `out' is 
purely formal.  

Now, we are going
to be assuming throughout this paper that the energy density 
$\sigma$ of any eight-brane
is positive, on the grounds that the tension should be proportional to some R-R 
charge density.  Following the analysis of \cite{ber}, it turns out that 
there are three basic cases, which are determined by the relative signs
of ${\epsilon}_{out}$ and ${\epsilon}_{in}$.  These cases are exhibited by
first introducing the parameter 
\[
{\Gamma} = \frac{{\Lambda}_{out} - {\Lambda}_{in}}{576{\pi}G{\sigma}^{2}}
\]

\noindent which
more or less measures the relative sizes of the `in' and `out' regions.
The cases can then be classified as shown in the below table:
\vspace*{0.3cm}

\begin{center}
\begin{tabular}{|c|c|c|c|}
\hline
 &Value &Value &Value  \\
Different &of &of &of \\
Cases &${\Gamma}$ &${\epsilon}_{out}$ &${\epsilon}_{in}$ \\
\hline
 & & &  \\
Case 1 &${\Gamma} > 1$ &+1 &+1 \\
 & & &  \\
\hline
 & & &  \\
Case 2 &$-1 < {\Gamma} < 1$ &-1 &+1 \\
 & & &  \\
\hline
 & & &  \\
Case 3 &${\Gamma} < -1$ &-1 &-1 \\
 & & &  \\
\hline
\end{tabular}
\end{center}
\begin{center}
{\bf Table 2}
\end{center}
\vspace*{0.6cm}

In Case 1, the in region is small relative to the out region, whereas
in Case 2 the regions are of comparable size.  Case 3 is equivalent to
Case 1, with the in and out regions interchanged.  In all of these cases the motion
of the eight-brane bubble, as seen by free-falling observers on either side,
is qualitatively similar.  Indeed, we completely analyzed this spacetime
in our analysis above.  For this class of spacetime, the brane is spherical
($k > 0$) and the cosmological constant is positive, so we know that
the scale factor $f(t)$ goes like
\[
f(t) = {\alpha}e^{ct} + {\beta}e^{-ct}
\]

\noindent where $k = 4{\alpha}{\beta}c^2$ ($\alpha$ and $\beta$ have the same sign)
and $t$ is the time experienced by observers who move with the
brane.  In other words, the brane expands exponentially with co-moving time.
Thus, in a sense the eight-brane `contributes' to the effects of inflation, 
because its positive tension generates a repulsive gravitational field.
This is very interesting, especially when we recall that these Dirac branes
are really just another type of supergravity domain wall \cite{mir}.
In \cite{mir} a complete classification was performed for the 
homogeneous and isotropic domain walls
which can arise in $N = 1$, $D = 4$ supergravity theory; there, the ground
state of the theory had {\it negative} cosmological constant.  Thus, in
$N = 1$, $D = 4$ supergravity domain walls are generically constructed
by pasting together sections of {\it anti}-de Sitter spacetime.
It was also shown that in a certain limit the
$D = 4$ domain walls are supersymmetric
(we will discuss the supersymmetric aspects of these Dirac branes in
more detail shortly).
In \cite{pairs} it was shown that black holes will generically be 
pair produced in the presence of repulsive, spherical domain walls, 
and indeed the arguments developed in those papers will go through here.
That is to say, ten-dimensional black hole pairs will be spontaneously
nucleated in the presence of the repulsive, de Sitter - de Sitter type
Dirac branes which we have been considering so far.  As we shall see, 
this is just one of the semi-classical instabilities which can affect
Dirac branes.

Now that we have considered the de Sitter - de Sitter type vacuum Dirac
branes, let's think about what happens when we turn the mass parameter $C$
back on.

We begin by considering the simplest situation where a Schwarzschild term
arises, namely, the situation illustrated in Figure 1 above.  In other words,
we have a bubble of de Sitter bounded from an (asymptotically flat) region
of Schwarzschild.  Thus, inside the bubble we have
\[
f_{in} = 1 - \frac{\Lambda}{36}r^2
\]
and outside
\[
f_{out} = 1 - \frac{C}{r^7}
\]

We insert these functions into (3.35), and then rearrange terms,
to obtain the below expression for the equation of motion:
\begin{equation}
\frac{C}{r^7} = r^{2}(\frac{\Lambda}{36} + \frac{{\kappa}^{2}{\sigma}^{2}}{4})
+ {\kappa}{\sigma}r{\sqrt {{\dot r}^2 + 1 -\frac{C}{r^7}}}
\end{equation}

\noindent Following \cite{guth}, let ${\zeta} = \frac{{\kappa}{\sigma}}{2}$, 
$\frac{\Lambda}{36} = {\xi}^2$, and $R^2 = {\zeta}^2 + {\xi}^2$, then after
a brief calculation (3.5) assumes the form
\begin{equation}
(\frac{dr}{d\tau})^2 + V(r) = E
\end{equation}

\noindent where the `potential energy' $V(r)$ is given as
\begin{equation}
V(r) = \frac{-C}{r^7} - \frac{C^2}{4{\zeta}^{2}r^{16}} + \frac{CR^2}{2{\zeta}^{2}r^7}
- \frac{R^{4}r^2}{4{\zeta}^2}
\end{equation}

\noindent and the total energy $E$ is $E = -1$.  As always, $\tau$ denotes 
co-moving time.  Thus, the motion of the eight-brane reduces to the motion of
a particle in a one-dimensional potential well.  It is not too hard to show
that the potential energy (3.38) generically has a unique maximum, at which
\[
\frac{dV}{dr} = 0
\]

\noindent For the edification of the reader, we include below a plot of
$V(r)$ for generic values of $M$, $\Lambda$, and $T$:
\vspace*{0.3cm}

\epsfxsize=10cm
\epsfysize=12cm
\rotate[r]{\epsfbox{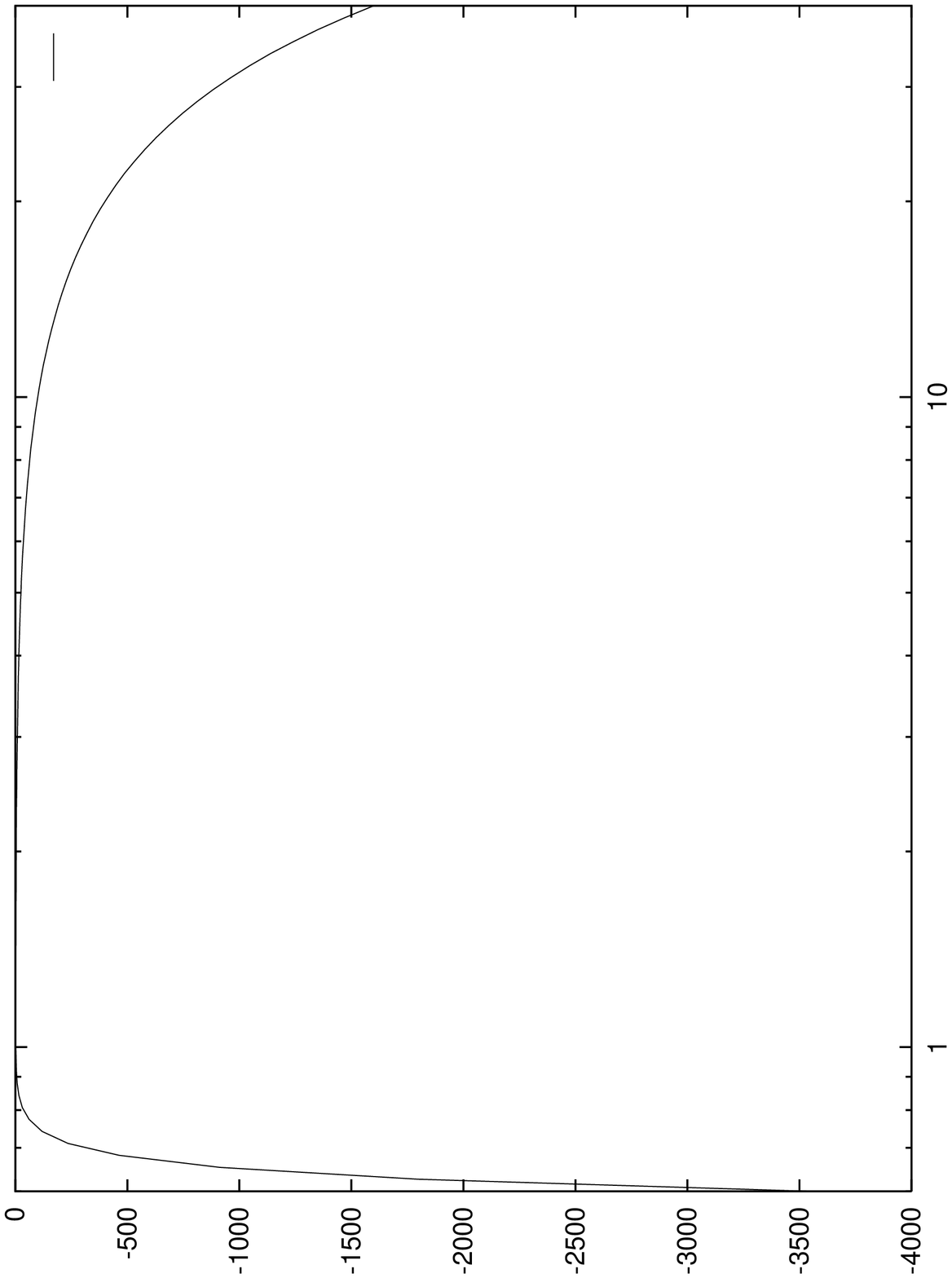}}
\vspace*{0.2cm}

It is clear, then, that in general there will exist a {\it unique} critical
radius $r_c$ at which the Dirac brane will be static.  It is also clear that
this radius will be unstable, i.e., generic perturbations of a static Dirac
brane will cause the brane to collapse or expand.  A rather tedious calculation
shows that $r_c$ is given explicitly as
\[
r_c = \Bigl[ \frac{(7/2)C[{\zeta}^2 - {\xi}^2] + 
C{\sqrt {9{\xi}^4 + 10{\xi}^{2}{\zeta}^2 + 44{\zeta}^4}}}{R^4} \Bigr]^{1/9}
\]

\noindent Thus, branes larger than the critical radius will want to expand,
and branes smaller than the critical radius will want to collapse.  It is not hard
to generalize this analysis to the situation where there is a mass parameter
{\it and} you are in a massive phase outside of the brane; in that situation,
the exterior of the brane is just Schwarzschild-de Sitter space, and the 
basic equations are very similar with some minor changes.

Of course, an interesting generalization is to turn on the (massive) $U(1)$ 
gauge field which comes from the dimensional reduction from eleven dimensions,
so that you are basically looking for spherically symmetric black hole solutions
in the Einstein-Maxwell-dilaton system described by (2.2)-(2.4).  
Naively, one might think that this would be important for understanding 
when these branes will bound supersymmetric vacua.  However, we will not
pursue further here the question of when these branes are SUSY, for the 
simple reason that there are some very subtle problems with the $N = 2$
SUSY algebra in ten dimensions when the cosmological constant
is positive.  We feel that it is better to postpone a thorough treatment
of this issue for a future paper, which will be devoted to this subject.

\section{Semiclassical instabilities of Dirac branes}

The nucleation and annihilation of bubbles of phase transition in the early
universe has been studied by a number of authors \cite{shawn1}, \cite{guth},
\cite{ber}, \cite{fgg}.  Here, we briefly sketch how these results will
also go through for Dirac branes.

In \cite{fgg}, the authors calculated the probability that one could
`create a universe in the laboratory' by quantum tunneling.  By this, they
meant ``what is the probability that a bubble of false vacuum (i.e.,
de Sitter space)  can appear in a lab where we have attained a super-high
mass density of the order $10^{76}$ g/$cm^{3}$ (i.e., the `lab' is a black
hole)''?  The authors calculated the rate at which one could create new
universes in this way by first finding an instanton, or imaginary time path, which
interpolates between the initial state (Schwarzschild) and the final state
(Schwarzschild with a bubble of de Sitter in it), then working out the
Euclidean action $S_E$ for the instanton path and then using the standard
semiclassical approximation for the probability $P$:
\[
P {\propto} e^{-S_{E}}
\]

\noindent Crudely, they found that the Euclidean action goes like
\[
S_E {\propto} 1/H^2
\]
\noindent where $H = (\frac{\Lambda}{3})^{1/2}$ is the inverse of the Hubble
radius.  This form of the action follows from the fact that the domain wall
sweeps out a three-sphere in imaginary time, and the action is basically the
wall tension times the volume of the three-sphere.

In ten dimensions, a Dirac eight-brane sweeps out a nine-sphere in imaginary
time; it follows that the Euclidean action for the nucleation of Dirac branes
will go like $1/H^8$.  In other words, the nucleation of isolated bubbles of massive
phase in IIA supergravity will still be highly suppressed.

This shows that Dirac branes can be spontaneously nucleated, but what about
the time reverse: How do they annihilate?  In a recent and interesting paper,
Kolitch and Eardley \cite{shawn1} studied the decay of  
Vilenkin-Ipser-Sikivie (VIS) domain walls in cosmology.  These VIS domain
walls are the Minkowski-Minkowski version of the de Sitter-de Sitter (dS-dS)
Dirac branes discussed above.  That is to say, a VIS domain wall is a repulsive
spherical bubble which seperates two (compact) portions of Minkowski space
(just as a dS-dS Dirac brane seperates two compact portions
of de Sitter spacetime).  It is not hard to see that their construction
will in fact go through for the dS-dS Dirac branes.  Again, the action for 
the instanton describing such an annihilation event will go crudely as
$1/H^8$.  

We have already mentioned that black hole pairs will be nucleated in the
presence of the repulsive, dS-dS spherical Dirac branes.  This is basically
because these dS-dS branes are bubbles bounding two regions of inflationary
phase, and we know that black holes will be produced in an inflating
(or domain wall) background.  Of course, just about {\it anything} can
be produced in an inflationary background, simply because
the repulsive gravitational energy is a natural source capable
of pulling virtual loops of matter out of the vacuum.
In particular, it is well known that topological defects \cite{review}
will also be nucleated in inflation.
Typically, a defect nuleates at the Hubble radius $r = H^{-1}$.
If the defect is much thinner than the scale
of the universe at the moment of nucleation, it makes sense to model the
defect using the Nambu action.  That is, it makes sense to assume that the 
defect is `infinitely thin', and to define the action to be the area of the
worldvolume swept out by the defect (multiplied by the characteristic tension,
or mass, of the defect).  In this limit, the instanton for a loop of 
string is a two-sphere of radius $H^{-1}$.  Similarly, the instanton for a 
(closed) spherical
domain wall is a three-sphere of radius $H^{-1}$, and so on.
Thus, the Euclidean action
for defect nucleation generically has the form
\[
S_{E} = {\mu}Vol(S^{n}(1/H))
\]
\noindent where $Vol(S^{n}(1/H))$ denotes the volume of an n-sphere of radius
$1/H$, and $\mu$ denotes the mass of the monopoles (if $n = 1$), the tension of
the string loop (if $n = 2$) or the energy density of the domain wall (if $n = 3$). 
One therefore finds that the rate of production of these defects is strongly
supressed if the defect tension is very large, or the cosmological constant
is very small, as would be expected.  Similarly, light defects are likely to
be produced in a background with a large cosmological constant.

The point is, there is no reason why these arguments will not apply
to the nucleation of loops of {\it fundamental} string in the presence
of the dS-dS Dirac branes.  This sort of string production is rather
reminiscent of crossing symmetry.  Indeed, fundamental string loops
will be created by this process, and it would be interesting to understand
the evolution of these strings after they appear.

It would also be interesting to compare the rate of brane annihilation to, say,
the rate at which black holes, string loops,
or other objects are nucleated in the presence of a brane.

\section{Conclusion}

Perhaps the most exciting thing which these results teach us is that
it is possible to describe gravitating brane configurations without losing
sight of the brane worldvolume.  To put it another way, there is nothing to
stop you from defining an effective worldvolume action for these gravitating
Dirac branes (presumably the Born-Infeld-Dirac action will do), and studying
these branes from the worldvolume point of view.

This should be contrasted to other `heavy', 
or gravitating, brane configurations in supergravity
theories, such as the super p-branes for instance \cite{gazpaul}.  There, the
brane worldvolume `goes away', and you are simply left with a geometry which looks
rather like a black hole (extended in some extra dimensions), such that the
solution `interpolates' between different vacua (generically Minkowski spacetime
`far' from the brane, and $(adS)_{p+2} {\times} S^{D - p -2}$ as you
get `near' the brane, as discussed in
\cite{gazpaul}).  In these solutions, there is no brane to be found, and 
therefore it is meaningless to talk about worldvolume actions.

Of course, given the recent results (\cite{curt}, \cite{gazz}) concerning
how a string ending on a D-brane will `tug' on the brane, this begs the 
question:  How will a `heavy' string tug on a heavy D-brane?  One has
to be careful in posing this question, since it is not at all clear that
the method of assigning distributional curvatures to spacetimes of low
differentiability (i.e., this is what we did when we
imposed the Israel matching conditions) will still make sense when one
is dealing with extended objects which are not domain walls, that is, where
the codimension of the worldvolume relative to the spacetime dimension is 
greater than one.
Research into these problems is currently underway.


{\noindent \bf Acknowledgements}\\

The authors would like to thank Gary Gibbons, Mike Green, Julius Kuti,
and Neil Lambert for useful
conversations.  A.C. is supported by a Drapers Research Fellowship at
Pembroke College, Cambridge.  M.J.P. was partially supported by a grant
from the U.S. Department of Energy.

\end{document}